# Advanced XR-Based 6-DOF Catheter Tracking System for Immersive Cardiac Intervention Training

Mohsen Annabestani[1], Sandhya Sriram[2,3,4], S. Chiu Wong[5], Alexandros Sigaras[2,3,4] and Bobak Mosadegh[1*]

1. Dalio Institute of Cardiovascular Imaging, Department of Radiology, Weill Cornell Medicine, NY, USA

2. Englander Institute for Precision Medicine, Department of Physiology and Biophysics, Weill Cornell Medicine, NY, USA

3. Institute for Computational Biomedicine, Department of Physiology and Biophysics, Weill Cornell Medicine, NY, USA

4. AI-XR Lab, Department of Physiology and Biophysics, Weill Cornell Medicine, NY, USA

5. Division of Cardiology, Department of Medicine, Weill Cornell Medicine, NY, USA

**Abstract:** Extended Reality (XR) technologies are gaining traction as effective tools for medical training and procedural guidance, particularly in complex cardiac interventions. This paper presents a novel system for real-time 3D tracking and visualization of intracardiac echocardiography (ICE) catheters, with precise measurement of the roll angle. A custom 3D-printed setup, featuring orthogonal cameras, captures biplane video of the catheter, while a specialized computer vision algorithm reconstructs its 3D trajectory, localizing the tip with sub-millimeter accuracy and tracking the roll angle in real-time. The system's data is integrated into an interactive Unity-based environment, rendered through the Meta Quest 3 XR headset, combining a dynamically tracked catheter with a patient-specific 3D heart model. This immersive environment allows the testing of the importance of 3D depth perception, in comparison to 2D projections, as a form of visualization in XR. Our experimental study, conducted using the ICE catheter with six participants, suggests that 3D visualization is not necessarily beneficial over 2D views offered by the XR system; although all cardiologists saw its utility for pre-operative training, planning, and intra-operative guidance. The proposed system qualitatively shows great promise in transforming catheter-based interventions, particularly ICE procedures, by improving visualization, interactivity, and skill development.

# 1. INTRODUCTION

Minimally invasive interventions (MII) have revolutionized the field of cardiac care, offering patients reduced recovery times, lower risks of complications, and shorter hospital stays compared to traditional open-heart surgeries. These procedures, such as percutaneous cardiac interventions, rely on the precise navigation of catheters through complex vascular structures and heart chambers[1-6]. However, traditional imaging techniques, including fluoroscopy and echocardiography, only provide limited information about the catheter's orientation and position, requiring clinicians to mentally reconstruct the catheter's path [7, 8]. As these procedures become increasingly complex, there is a growing need for advanced systems that offer enhanced visualization and guidance, particularly with comprehensive six degrees of freedom (6-DOF) tracking[9].

The concept of 6-DOF refers to an object's ability to move freely in three-dimensional space. It encompasses three translational movements (up/down, left/right, forward/backward) and three rotational movements (pitch, yaw, and roll) [9, 10]. In the context of cardiac catheterization, 6-DOF tracking is crucial for precise navigation and manipulation, especially for specialized tools like ICE catheters. While all six degrees of freedom are important, the roll angle – rotation around the catheter's longitudinal axis – plays a particularly critical role in many procedures[9]. It enables precise navigation by allowing the catheter to be rotated to access different parts of the heart and avoid obstacles. Furthermore, it enhances visualization by changing the view direction of imaging catheters, crucial for procedures using ICE catheters. Precise roll control also increases procedural flexibility, allowing operators to adapt to patient-specific anatomical variations, and can significantly reduce procedure time by facilitating quick adjustments to obtain optimal imaging views[11, 12]. In complex procedures like transcatheter aortic valve implantation (TAVI) and transcatheter mitral valve replacement (TMVR), precise roll control is essential for accurate device placement [13-16].

Despite the crucial role of roll angle information in catheter procedures, existing imaging and tracking technologies often fail to deliver accurate, real-time data in this area. This gap presents a major hurdle in both clinical applications and training environments. To overcome this, we have developed a 3D-printed electromechanical encoder for real-time roll angle measurement of ICE catheters, alongside a cutting-edge extended reality (XR) system that offers complete 6-DOF catheter tracking and visualization. XR headsets enable users to view digital content superimposed on the real world, augmenting their natural view with instructional 2D and 3D content. These advancements are highly relevant for a range of applications, including percutaneous transcatheter cardiac intervention training [12], pre-procedural planning, intraoperative guidance [11], and telesurgery, where remote catheter operation is required [17]. Our approach delivers several advantages over traditional methods, providing comprehensive real-time feedback on catheter orientation and precise roll angle tracking. The system also integrates with XR technology, creating an immersive and user-friendly interface for both training and procedural support. Additionally, it addresses the shortcomings of 2D fluoroscopic imaging by enhancing depth perception and spatial awareness. The applications of this technology are far-reaching. In the realm of medical training, it offers a safer, more cost-effective alternative to traditional fluoroscopy-guided practices [18-20]. The system's ability to replicate real procedural tools within an XR environment allows trainees to develop psychomotor skills that closely mimic live cases[21-23].

The results of this research extend beyond cardiac interventions, offering significant insights for the broader field of medical training. By advancing the understanding of the advantages and limitations of 3D visualization in XR-based systems, this work contributes to the ongoing efforts to refine and enhance simulation technologies. Our development introduces an innovative approach to XR-based cardiac catheterization, featuring the ability to measure and visualize the roll angle. Utilizing the Meta Quest 3 and real-time 3D catheter reconstruction, which mirrors physical manipulation by the user, this system creates an accurate 3D model of the catheter within a patient-specific heart rendering. This precise integration within the Meta Quest 3 provides improved visualization and a more immersive training experience compared to traditional techniques. The multi-view perspectives offered by the system have the potential to reduce the learning curve for mastering new catheter systems and to better equip practitioners for handling cases involving complex anatomies.

This paper is organized into three main sections. In Part 2, we will outline our methodology for the proposed system and provide detailed descriptions of both the hardware and software components of the XR system. Section 3 will present the results and findings of the study. Finally, in the conclusion, we will summarize the key outcomes of our work.

## 2. METHODOLOGY

In the proposed method, a novel real-time 6-DOF catheter tracking system has been developed, designed to be fully compatible with the Meta Quest 3 XR headset. The system is capable of working with a wide range of commercially available catheters and consists of two primary hardware components: a 3D-printed vision box and an electromechanical encoder. The 3D-printed vision box is equipped with two orthogonally positioned cameras that capture biplane images of the catheter (**Figure 1-b,c**). A custom computer vision algorithm analyzes these images to determine the catheter's orientation and shape, which is then used to reconstruct its 3D configuration (**Figure 1- d, e, f, g**) and extracting real-time 5-DOF features. The electromechanical encoder (**Figure 1-h**), in conjunction with an Arduino board (**Figure 1-i**) and a custom Python program (**Figure 1-j**), enables real-time measurement of the roll angle, thereby completing the 6-DOF tracking.

Real-time catheter data, including the roll angle, is transmitted to the XR headset using the WebSocket protocol (**Figure 1-k**). Within this setup, the Unity game engine drives an independent rendering environment (**Figure 1-n**). This environment displays a reconstructed catheter, and visualized roll angle, within a patient-specific heart model (**Figure 1-m**), providing users with a highly immersive and interactive experience. The heart model is generated from a cardiac CT scan captured in DICOM format at end-diastole (**Figure 1-l**). The catheter data is co-registered with this anatomical representation within the Meta Quest 3 (**Figure 1-o**), allowing users (**Figure 1-p**) to manipulate a real commercial catheter and observe its real-time movement within the 3D heart model. The subsequent sections offer a comprehensive technical overview of each stage of development.

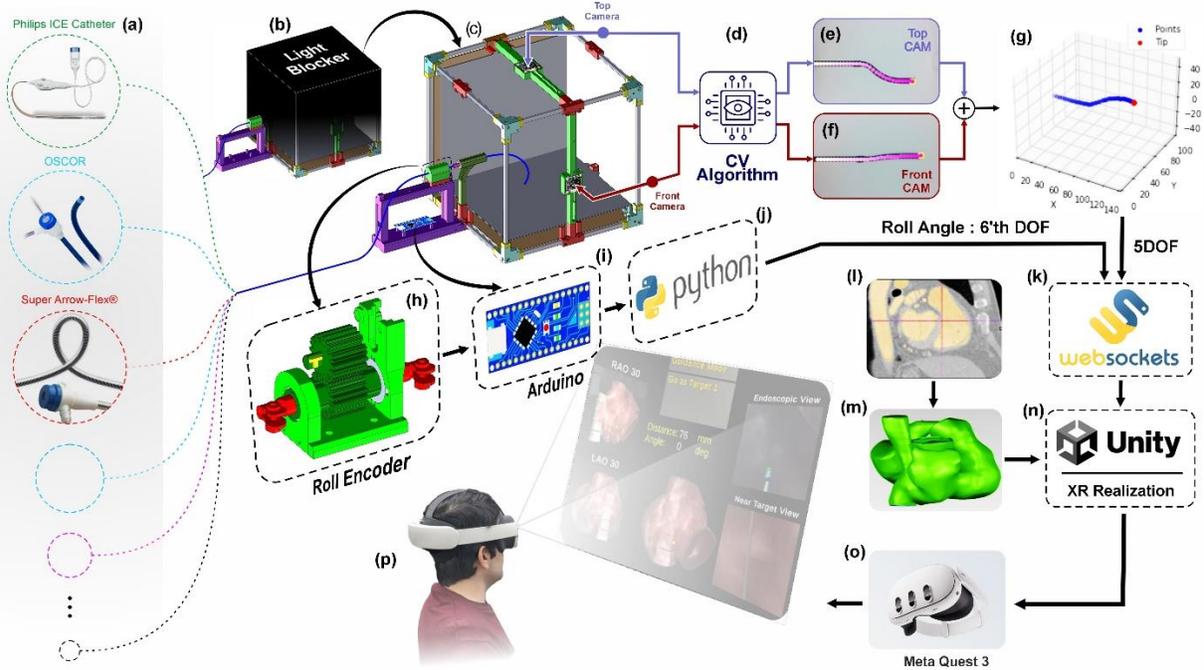

**Figure 1:** The proposed XR-based system: (a) commercially available catheters utilized in the system; (b) vision box with a light blocker; (c) vision box without a light blocker; (d) custom computer vision algorithm; (e & f) inferred catheter shapes derived from biplane views; (g) the reconstructed 3D shape of the catheter; (h) the encoder designed for measuring the roll angle of ICE catheters; (i) an Arduino board for recording and transmitting encoder values; (j) a Python program that converts encoder values into roll angles; (k) a WebSocket unit; (l) segmented blood volume from a de-identified CT scan (DICOM); (m) a patient-specific 3D heart mesh model; (n) a Unity-powered rendering environment; (o) a Meta Quest 3 XR headset; and (p) a user manipulating a real catheter within the 3D model of a patient's heart while observing the process through the headset.

## 2.1. 3D-Printed Vision Box

To enable catheter tracking and provide a stable platform for controlled manipulation, we developed a cubic 3D model using Dassault Systems SolidWorks 2022 and fabricated it with a FDM 3D printer. As shown in **Figure 2**, several key elements were incorporated into the setup to ensure accurate computer vision tracking. A removable inlet, positioned at the origin (0,0,0) of the 3D Cartesian coordinate system, allows catheter insertion and movement within the cube's central open space. This inlet is interchangeable to accommodate catheters of different diameters. Additionally, mounts were integrated into the design to securely hold two cameras in orthogonal positions along two sides of the cube, forming a biplane imaging system. These cameras capture video of the catheter as it moves through the central region of interest (ROI).

To define the ROI, we positioned eight fiducial markers (mTi and mFi for i=1,2,3, and 4, where T and F stand for Top and Front) at specific points on two cross-shaped pillars within the model. These markers ensure that the ROI is appropriately sized for a human heart. The cross-shaped pillars, used only temporarily for marker detection, are removable once the markers have been identified. The fiducial markers are placed on orthogonal planes, allowing for the transformation of each camera's viewpoint into a global coordinate system. This coordinate transformation is vital for reconstructing the catheter's 3D trajectory and ensuring proper alignment between the heart model and the catheter inlet.

To eliminate external light interference and fully control lighting conditions, we enclosed the vision box with light-blocking layers and incorporated internal LED lighting systems, coupled with a PMMA light diffuser for background illumination. The LED planes, laser-cut from translucent PMMA sheets, provide a high-contrast background to enhance the visibility of the catheter for the cameras, regardless of its color or type. This design helps the computer vision algorithm to accurately isolate and segment the catheter. Except for the connecting pillars (as shown in **Figure 2**), the entire setup was 3D printed from PLA thermoplastic using a PRUSA MK4 FDM 3D printer. The components are modular and connected using neodymium magnets for easy assembly and disassembly.

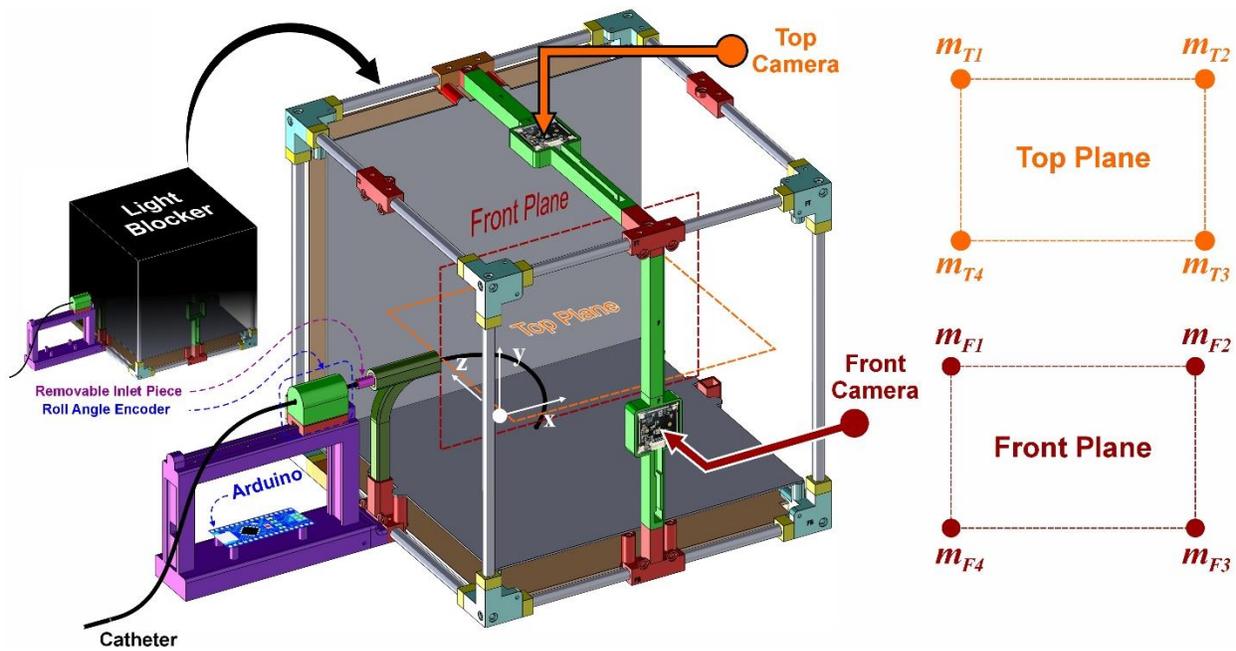

**Figure 2:** The schematic design of the proposed Vision Box includes a removable inlet fixture for catheter insertion at the center of the region of interest (ROI) and a mount for the Roll Angle encoder positioned on a linear rail. The box features four color-coded reference points (mTi and mFi, where i=1,2,3, and 4, with T and F denoting Top and Front), which define the cubic boundary of the heart. Additionally, two mounts are placed on opposite sides of the cube to securely hold two cameras, creating an orthogonal biplane imaging system.

## 2.2. Machine Vision-Based Algorithm for 5DOF Catheter Tracking

In this paper, we present an upgraded computer vision algorithm for 5DOF catheter tracking, enhancing the robustness and speed of our previously developed algorithm [24].The new algorithm incorporates several submodules that effectively increase accuracy and speed, remove artifacts and fill gaps in the segmented catheter areas, addressing limitations in the previous version. The core of our tracking system continues to rely on a segmentation algorithm that identifies and segments the catheter from synchronized orthogonal camera footage. This pipeline processes video streams in real-time from two intrinsically calibrated cameras positioned around the 3D workspace. Each frame undergoes a perspective transformation, utilizing known fiducial points to create a unified coordinate space. After the perspective transformation, preprocessing steps such as Gaussian smoothing and contrast adjustments enhance the visibility of the catheter. An adaptive thresholding operation, which computes a dynamic local threshold for each pixel, further isolates the catheter.

A significant improvement in the new algorithm is the introduction of Breadth-First Search (BFS) for the tip-to-tail block, replacing the previous method. This change enables a more efficient and accurate reconstruction of the catheter's trajectory. Breadth-First Search (BFS) is a graph traversal algorithm that explores all neighboring nodes at the current depth before moving on to the next level[25]. In our context, BFS is utilized to find the optimal path between the tip of the catheter and its end, treating the skeleton pixels of the catheter as nodes in a network. This allows for efficient navigation through the catheter's shape, ensuring accurate tracking of its trajectory. This process creates a two-dimensional directional array that captures the spatial path from the tip to the entry point. These coordinates are mapped to a real-world 3D system in millimeters, utilizing known dimensions and camera intrinsic parameters.

The catheter's trajectory is represented by a down sampled set of K coordinate points. The first of these points corresponds to the catheter's distal tip, while the last point indicates its entry location. Between these, K-2 points are uniformly spaced to capture the catheter's curvature accurately. Our algorithm, developed in Python with OpenCV and custom-developed libraries, processes the orthogonal camera feeds in real-time. The algorithm extracts Y coordinates from the top-view camera and Z coordinates from the front-view camera, while X coordinates are derived from a combination of both top and front views. By synthesizing these datasets, we achieve a dynamic 3D K-point tracking mechanism. This approach yields a comprehensive, real-time visualization of the catheter's three-dimensional structure and bending profile.

## 2.3. Roll Angle Encoder for Required 6$^{th}$ DOF of ICE Catheter

We designed and implemented an electromechanical encoder system to measure the roll angle of an ICE catheter, as illustrated in **Figure 3**. The system comprises a gearbox, including two gears, where the bottom gear securely holds the ICE catheter using two soft TPU locks, ensuring a firm grip without damaging the catheter. When the catheter rotates, the bottom gear transfers this rotation to the top gear, which is directly connected to a rotary encoder. The rotary encoder, in turn, sends real-time signals to an Arduino Nano, which communicates with a Python script to calculate the catheter's roll angle in real time.

The entire setup, including the gearbox and rotary encoder, is mounted on a base that slides along a linear rail, ensuring that the ICE catheter's locking mechanism does not interfere with its linear movement into the vision box. All components — the gearbox, encoder, base, and rail — form a modular system that can be easily attached to the vision box when tracking the ICE catheter.

To ensure proper alignment, the system is designed to keep the catheter centered with the vision box inlet. During setup, we designate the right side of the vision box as the zero-degree roll position. A custom jig holds the catheter's transducer securely, ensuring accurate angle calibration. When the gear system is unlocked, the catheter can freely slide along the rail into the gearbox, allowing it to reach the end boundary of the vision box where the transducer jig is located. Once the ICE transducer is placed in the jig, the gear's transducer marker, labeled "T", should be aligned with the right side of the box, after which the catheter should be locked in place. This setup ensures the ICE catheter's zero-degree roll position is properly defined, allowing for smooth movement along the rail to the heart model's boundary.

The Arduino Nano is programmed once using Arduino language, and further processing, including roll angle calculation and automatic COM port detection, occurs within Python. For real-time data acquisition, the system operates with a baud rate of 115200. The system was modeled in SolidWorks and fabricated using a PRUSA MK4 3D printer with PLA thermoplastic material. This modular, easily attachable system ensures precise, real-time roll angle measurement without impeding the catheter's movement, providing a practical solution for real-time tracking of ICE catheters.

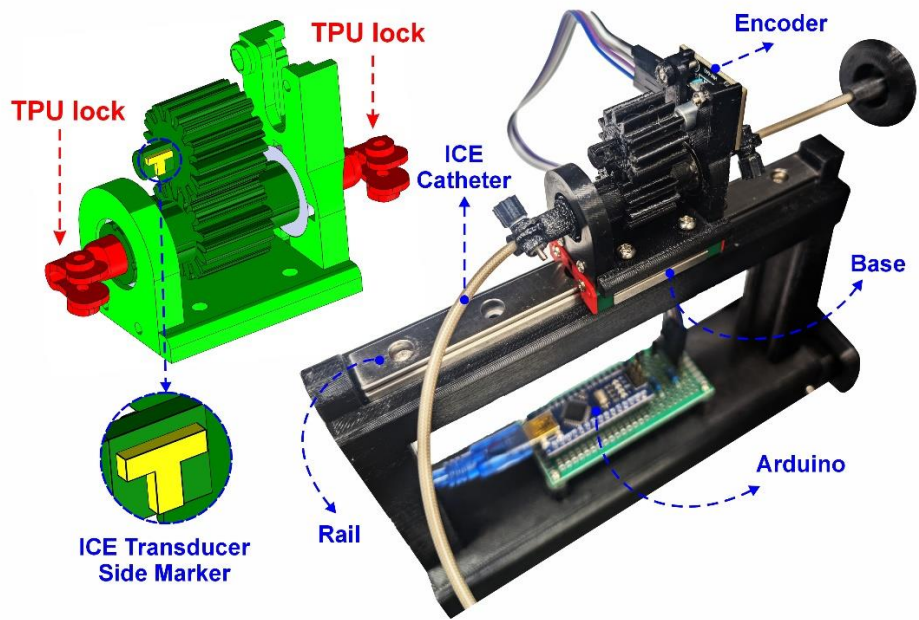

**Figure 3:** Proposed Roll Angle Encoder.

## 2.4. Extended Reality (XR) Rendering

Extended Reality (XR) is employed to integrate real-time catheter tracking with a patient-specific heart model, providing an immersive visualization experience. This XR environment, rendered using the Unity game engine on the Meta Quest 3 headset, combines the spatial configuration of the catheter with an accurate 3D representation of the patient's heart. The heart model is generated from a cardiac CT scan, processed and refined through Materialize Mimics and Geomagic Wrap software, as explained in our previous work [24].

The catheter's 3D position, derived from the real-time K-point data output by our computer vision algorithm, is rendered in the XR scene alongside the heart model. The catheter is rendered as a spline object in Unity, with knots on the spline being dynamically added and modified depending on the K-point data to form the shape and curve of the catheter. Communication between the Python-based tracking system and Unity game engine is facilitated via Flask, using a WebSocket connection to ensure real-time data transmission. The user connects the system by entering an IP address and port, allowing dynamic updates to the catheter's position and curvature within the virtual heart.

A crucial feature of the system is the visualization of the ultrasound beam emitted from the tip of the ICE catheter. This beam is represented as a trapezoidal-shaped field extending from the catheter's tip (**Figure 4**), providing real-time feedback on the roll angle of the catheter. The visualization of the ultrasound beam helps trainees and surgeons accurately align the ICE catheter transducer, ensuring proper orientation and positioning within the specific target in the heart. This is particularly valuable for guiding procedures, as the roll angle directly influences the direction of the beam, helping users locate the appropriate anatomical structures. Additionally, a spherical marker is placed within the scene (at the end of ultrasound beam) as a reference point for reaching the targets, allowing users to practice precise catheter placement and alignment. This visual guidance helps the user to locate the transducer in the optimal position, enhancing both the training and real-life procedural accuracy by providing clear visual feedback on catheter movement and positioning.

For accurate mapping, fiducial markers in the 3D heart model are used as reference points, enabling an affine transformation that aligns the catheter's movements with the heart's anatomical structure. This ensures that the catheter is correctly positioned within the heart model, creating a cohesive and interactive XR experience. The system provides quantitative feedback on catheter movements including distance and angle differences from the target, allowing trainees to practice complex catheterization procedures with real-time guidance, making it a valuable tool for both training simulations and potential clinical applications.

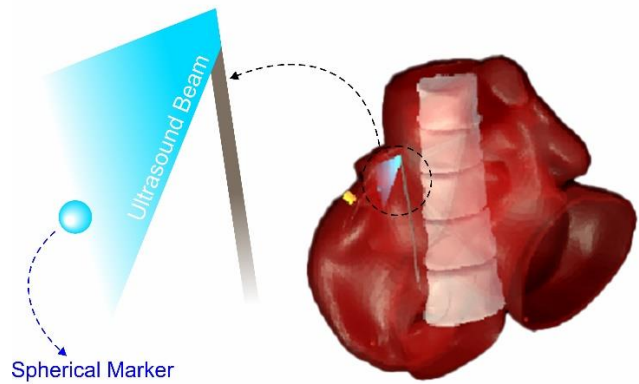

**Figure 4:** The visualization of the ultrasound beam emitted from the tip of the ICE catheter

## 3. RESULTS AND DISCUSSION

In this study, we present several improvements to our previously developed XR-based catheter tracking system [24], with enhancements in functionality, visualization, and system accuracy. One of the key improvements is the addition of an ICE catheter mode, accessible from the settings menu, which offers specific features tailored to ICE procedures. In ICE mode, a trapezoidal-shaped ultrasound beam is visualized at the tip of the catheter, representing the field of view of the ultrasound transducer. This ultrasound beam rotates in real-time based on the measured roll angle provided by the roll encoder we developed, allowing users to easily visualize and understand the catheter's roll during procedures. Moreover, the endoscopic view has been replaced with a new display that shows the ICE catheter's ultrasound view, providing more relevant feedback for ICE-specific applications, as shown in **Figure 5**.

Additionally, we evaluated the accuracy of our system in reconstructing the 3D trajectory of the catheter. Similar to our previous work [24]**,** we once again demonstrated the system's high precision in tip localization, with an average error of less than 1 mm. The system also showed excellent fidelity in reconstructing the full 3D catheter trajectory, further validating the accuracy and robustness of the upgraded machine vision algorithm in tracking the ICE catheter's shape and spatial configuration in real time.

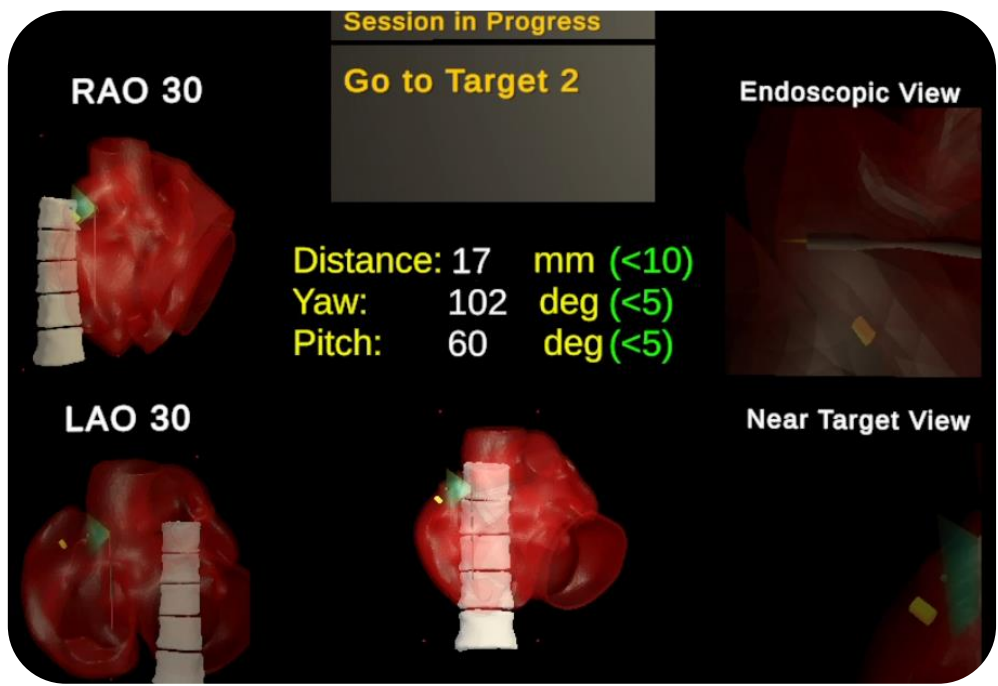

**Figure 5.** A sample scene of the proposed XR-based 6-DOFcatheter tracking system.

### 3.1 Proposed XR System Functionality and User Experience Assessment

To assess the functionality and user experience of the proposed XR system, we conducted a study involving six participants; it should be noted that this was only for system optimization and no generalizable conclusions can be made with this small sample size, and no results were shown to be statistically significant. These participants, a mix of cardiologists, interventionalists, and biomedical engineers and scientists, had varying levels of experience with delivery catheters. Importantly, none were authors of this paper. The main objective was to evaluate both the quantitative performance and qualitative feedback to determine the effectiveness of different visualization modes in navigating a catheter to predetermined targets within the heart, with a particular focus on understanding the impact of 3D depth perception in localizing an ICE catheter.

The study began with a structured familiarization session, allowing participants to become comfortable with both the XR system and the catheter. This session was intended to provide everyone a baseline understanding of how the system operated before starting the main experiment. Participants then performed two different tasks, each focusing on a specific visualization mode:

**2D View:** In this session, participants were shown two-dimensional views, similar to those seen on traditional 2D monitors (**Figure 6, left**). Their task was to guide the catheter to predetermined targets without depth perception, requiring mental coordination of multiple 2D views to estimate 3D positioning.

**3D View:** In this session, participants relied exclusively on a three-dimensional representation of the heart (**Figure 6, right**). The goal here was to see if the added depth perception from the 3D view improved performance over the 2D views.

To minimize bias from task order, all participants completed the 2D session first, followed by the 3D session. This ensured that any learning effects were consistent across both tasks. During each session, key performance metrics were recorded:

**Completion Time (t):** The time (in seconds) taken by each participant to complete the task.

**Number of Targets Reached (nT):** How many of the predetermined targets were successfully reached by the participants.

**Time per Target (tT):** The average time it took to reach each individual target, calculated as the total time divided by the number of targets reached.

The results of all six participants are summarized in **Figure 7**. On average, the participants completed the task **4.4% slower** using the 3D view compared to the 2D view. However, when we exclude one specific participant (Subject S3), this result changes dramatically to a **27% faster** performance in 3D. This subject performed much slower in the 3D view, making them an outlier. This discrepancy could be due to S3's background as a structural interventional cardiologist, who is highly accustomed to working with 2D views in the cardiac catheterization lab. In contrast, most other participants had little to no prior experience with catheters or the catheterization lab environment, with the exception of the electrophysiologist but they routinely use 3D visualization in electroanatomic mapping systems. Given S3's extensive experience with just 2D views, it's likely that they were more efficient in the 2D session, unlike the other participants, who benefited more from the additional depth perception provided by the 3D view. To reduce the impact of such outliers, future studies should include a larger sample size or have each participant complete multiple trials per session, using either the best or the average score to provide a more balanced view of performance.

Interestingly, in our previous study using another catheter type (OSCOR) [24], we found that the 2D view was 3x longer than the 3D view; it should be noted that this previous study was also only for system optimization and no generalizable conclusions can be made with the small sample size of 6 subjects. However, when considering a similar of population of non-physicians only, this study with the ICE catheter showed the 2D view was only 1.4x longer than the 3D view. However, given that only 2/3 non-physician subjects performed better in 3D vs 2D, this comparison is not strong. Also, given that the average 3D view durations were similar between these two studies, it suggests that the difference was due to the difficulty in maneuvering the OSCOR catheter in the 2D view. This can be potentially explained by the fact that the OSCOR only has 1 bending know while the ICE catheter has 2 bending knobs giving more degrees of freedom for movement. Given that not all users show a clear preference for the 3D view, it is important that these navigation systems provide options for users to accommodate their preferences and prior training. Future work will explore these dynamics further with an expanded sample, repeated trials, and more performance metrics (e.g., total path length, perforations).

For the qualitative assessment, we created a 10-question survey designed to gather insights into different facets of the participants' experiences. These qualitative findings reveal valuable insights into the participants' experiences with the extended reality image-guidance system for catheter-based procedures. The results of this survey are shown in Table 1. Despite the varied levels of experience among participants in interventional procedures, ICE, and video games, the system generally received positive feedback (>4 out of 5 on average). The onboarding training was received by most participants, with 3.6 out of five, suggests that a better onboarding process is needed for participants to feel more prepared for the experience. The 3D view mode, compared to the 2D view mode, was predominantly perceived as beneficial for improving accuracy in ICE catheter placement, with all participants rating it 3 or higher. However, its impact on improving speed showed more varied results, suggesting potential areas for further refinement.

The extended reality system demonstrated strong potential in both pre-operative and intra-operative applications, with all participants rating its usefulness as 4 or 5 out of 5 for these purposes. The digital interface was generally considered user-friendly, receiving ratings of 3 or higher from all participants, but likely needs some improvement. Notably, the overall experience with the system was largely positive, with four out of five participants rating it 4 or 5. These findings suggest that the extended reality image-guidance system shows promise as a valuable tool in catheter-based procedures, particularly in terms of accuracy improvement and potential applications in both training and live procedures. However, the varied responses regarding speed improvement indicate an area for potential enhancement in future iterations of the system.

Table 1. The results of qualitative survey

| Question # | Question | Subjects Rating: 1-5 (5 being the highest) | | | | | | |
| --- | --- | --- | --- | --- | --- | --- | --- | --- |
| | | S1 | S2 | S3 | S4 | S5 | S6 | Average |
| 1 | How much experience do you consider having with performing interventional procedures (i.e., catheter-based), relative to an experienced attending? | 2 | 5 | 5 | 1 | 1 | 1 | 2.5 ± 1.97 |
| 2 | How much experience do you consider having with performing intracardiac echocardiography (ICE), relative to an experienced attending? | 3 | 5 | 5 | 1 | 1 | 1 | 2.66± 1.97 |

| 3 | How well do you consider your on-boarding training for use of a catheter before performing the mock intervention? | 3 | 3 | 5 | 1 | 5 | 5 | 3.6 ± 1.63 |
|---|---|---|---|---|---|---|---|---|
| 4 | How helpful was the 3D view mode compared to the 2D view mode for improving accuracy of the ICE catheter placement to a target? | 3 | 5 | 3 | 3 | 5 | 3 | 3.6 ± 1.03 |
| 5 | How helpful was the 3D view mode compared to the 2D view mode for improving speed of finding a target position? | 3 | 5 | 2 | 2 | 5 | 3 | 3.33 ± 1.36 |
| 6 | How user-friendly was the digital interface of the extended reality image-guidance system? | 4 | 3 | 5 | 3 | 5 | 5 | 4.17 ± 0.98 |
| 7 | How useful would this extended reality image-guidance system be for pre-operative training/planning? | 4 | 4 | 4 | 4 | 5 | 4 | 4.17 ± 0.4 |
| 8 | How useful would this extended reality image-guidance system be for intra-operative procedures? | 4 | 5 | 4 | 4 | 5 | 4 | 4.33 ± 0.5 |
| 9 | In general, how was your overall experience with this system? | 4 | 4 | 5 | 3 | 5 | 5 | 4.33 ± 0.8 |
| 10 | How much experience do you consider having with video games? | 2 | 5 | 3 | 1 | 4 | 1 | 2.66 ± 1.63 |

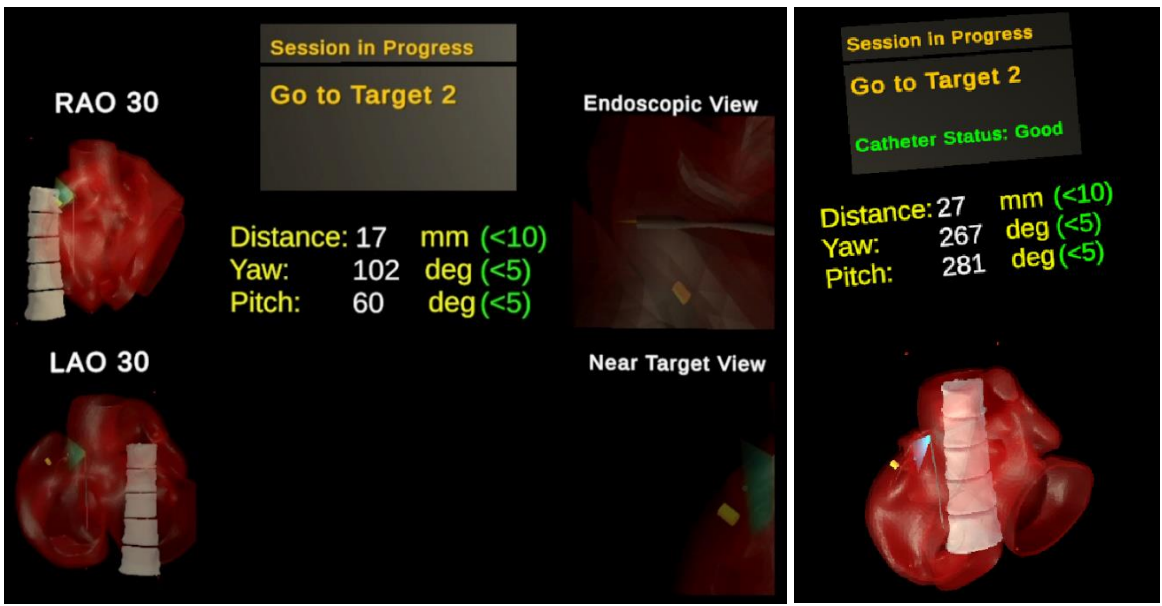

**Figure 6. Left:** 2D-View, **Right:** 3D-View

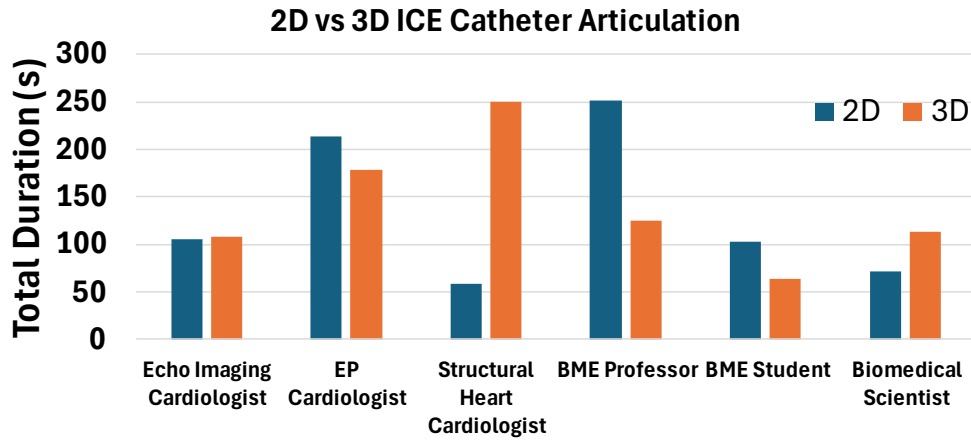

**Figure 7.** 2D vs. 3D ICE Catheter Articulation Results: Total time taken to reach all six targets in both the 2D and 3D visualization modes.

## 4. CONCLUSION

In this study, we developed an advanced XR-based 6-DOF catheter tracking system designed to enhance both medical training and procedural guidance during percutaneous cardiac interventions. By combining real-time catheter tracking with precise roll angle measurement and immersive XR visualization, our system addresses critical challenges in catheter navigation, especially for ICE catheter-based procedures. The experimental results and user feedback highlight the system's ability to improve depth perception, spatial awareness, and procedural accuracy, particularly when transitioning from traditional 2D views to 3D environments. Although there appears to be personal preferences as to whether a 2D or 3D view is better for navigation, it seems that the system overall will benefit all users in performing training and real-time guidance in cardiac interventions, which can be validated in future studies that compare results with and without this system. Future work will focus on expanding the user base, improving onboarding and performance metrics, and exploring its benefit for providing better clinical outcomes through pre-procedural training and planning.

## CONTRIBUTION

Methodology and Machine vision algorithm design, M.A.; Coding and implementation of the algorithm, M.A.; Design and implementation of the 3D setup, M.A.; Design, Fabricate, program and implementation of the Roll Encoder, M.A.; Testing and investigation, M.A.; Experimental study , and Data analysis, M.A.; Front-end development and coding, S.S.; Unity implementation and mixed reality, S.S.; Testing and investigation, S.S.; Resources, S.C.W.; Front-end project supervision, A.S.; Conceptualization, B.M.; Project administration, B.M.; Supervision, B.M.; Funding acquisition, B.M. All authors have read and agreed to the published version of the manuscript.

# Acknowledgement

We thank Philips for providing the VeriSight ICE catheter supported by a research grant. We also would like to acknowledge Alexandre Caprio for the development of the 3D heart model used in this study, which was created during a previous project and continues to contribute to our ongoing research.